\documentclass[showpacs,prb,preprint]{revtex4}
\usepackage{amssymb}
\usepackage{graphicx}

\begin{document}

\title{Non-volatile bistability effect based on electrically controlled
phase transition in scaled magnetic semiconductor nanostructures}
\author{Y. G. Semenov and K. W. Kim}
\address{Department of Electrical Computer Engineering,
North Carolina State University, Raleigh, NC 27695-7911}

\begin{abstract}
We explore the bistability effect in a dimensionally scaled
semiconductor nanostruncture consisting of a diluted magnetic
semiconductor quantum dot (QD) and a reservoir of itinerant holes
separated by a barrier.   The bistability stems from the magnetic
phase transition in the QD mediated by the changes in the hole
population.  Our calculation shows that when properly designed,
the thermodynamic equilibrium of the scaled structure can be
achieved at two different configurations; i.e., the one with the
QD in a ferromagnetic state with a sufficient number of holes and
the other with the depopulated QD in a paramagnetic state.
Subsequently, the parameter window suitable for this bistability
formation is discussed along with the the conditions for the
maximum robustness/non-volatility.  To examine the issue of
scaling, an estimation of the bistabiity lifetime is made by
considering the thermal fluctuation in the QD hole population via
the spontaneous transitions. A numerical evaluation is carried out
for a typical carrier-mediated magnetic semiconductor (e.g.,
GaMnAs) as well as for a hypothetical case of high Curie
temperature for potential room temperature operation.
\end{abstract}



\maketitle

The magnetism in semiconductors is the basis for the emerging
field of spin-polarized electronics, or spintronics.~\cite{Awsch}
Substantial progress has been made during the past few years,
particularly in the materials development. The advantages of the
semiconductor-based systems over the metallic counterparts include
the controllability of the ferromagnetism (via the
bias~\cite{Ohno} and/or doping) and the potential compatibility
with the modern Si-based processing technology.

Recently, a theoretical study further explored the opportunities
offered by the electrically controlled magnetism. The calculation
found that a properly designed structure consisting of magnetic
and non-magnetic semiconductor quantum wells (QWs) can exhibit the
bistability with respect to the paramagnetic-ferromagnetic (PM-FM)
phase transition when the process is controlled by the itinerant
carriers (holes).~\cite{SHK} This bistability effect in the
\emph{two-dimensional} system was predicted to persist even at
temperatures nearly as high as the critical temperature $T_{c}$ of
the PM-FM phase transition. Subsequently, a non-volatile memory
application was suggested citing the successful growth of
transition metal doped
semiconductors that are FM at or above room temperature (e.g., the nitrides,~%
\cite{Bedair} Ge,~\cite{Tsui} as well as some
II-VI's~\cite{Saito,Norberg}).

For practical realization of the proposed device application, it is highly
desirable to reduce the size of the magnetic layer (i.e., the active part of
the memory) without the loss of high temperature operability and
non-volatility. Consequently, a magnetic semiconductor quantum dot (QD) that
can exchange itinerant holes with a reservoir provides an interesting
opportunity. However, unlike the QW case examined earlier,~\cite{SHK} the
thermal fluctuation resulting from spontaneous hopping between the two
possible stable states cannot be neglected in the scaled structure due to
the limited number of carriers populating the QD. Hence, it is the purpose
of this paper to investigate theoretically the effect of the reduced
dimensionality and explore the potential bistability conditions based on the
electrically controlled magnetic phase transition in the magnetic
semiconductor nanostructures.

The specific structure under consideration consists of a single
diluted magnetic semiconductor (DMS) QD separated from a large
reservoir of itinerant holes, which controls the chemical
potential $\mu _{0}$ of the system. For simplicity, we assume $\mu
_{0}\gg k_B T$ and ignore the possible temperature dependence of
$\mu _{0}$. A non-magnetic QW filled with itinerant holes (for
example, through the modulation doping, etc.) can be used as the
desired reservoir (see Fig.~1).

To accurately describe this system, one needs to know the energy
structure of the QD that is a function of multiple parameters.
Particularly, the magnetic interactions must be taken into account
that leads to the PM-FM phase transition in the DMS QD. Note that
the analysis of different magnetic phase transition mechanisms is
beyond the scope of the present study (see, for details,
Ref.~\onlinecite{PRetal}). Instead, we assume that the main
magnetic properties [such as the critical temperature $T_{c}$ and
its dependence on the hole concentration, and the magnetization
dependence on temperature $M=M(T)$] can be obtained from the
measurements of the relevant DMS. This allows the use of a
semi-phenomenological approach to calculate the free energy of the
system.

In the present model, we approximate the free energy of the QD as
the sum of two terms, the magnetic ($F_{M}$) and non-magnetic
($F_{N}$) contributions. If the DMS QD is not far from the PM-FM
transition, the Landau expansion over the magnetization $M$ can be
applied for $F_{M}$:
\begin{equation}
F_{M}=-a(T_{c}-T)M^{2}+bM^{4}.  \label{f1}
\end{equation}%
The parameters $a$, $b$ and $T_{c}$ are the functions of hole population $j$
in the QD; particularly, the dependence $T_{c}=T_{c}(j)$ plays a crucial
role in the magnetic instability.~\cite{Guinea} In addition, $a$ and $b$ can
be expressed in terms of the fundamental properties of the magnet: The
Curie-Weiss law for magnetic susceptibility $\chi =C_{0}/(T-T_{c})$ at $%
T>T_{c}$ defines $a=1/4C_{0}$, while spontaneous magnetization $M_{s}=M_{0}
\sqrt{1-T/T_{c}}$ at $T<T_{c}$ provides $b=aT_{c}/2M_{0}^{2}$. Since Eq.~(%
\ref{f1}) is assumed to fully describe the magnetization of the
DMS QD, the thermodynamically stable state corresponds to the
specific magnitude of $M$ that gives rise to the free energy
minimum
\begin{equation}
F_{M}=-T_{c}\frac{M_{0}^{2}}{C_{0}}\left( 1-\frac{T}{T_{c}}\right)
^{2}\theta \left( 1-\frac{T}{T_{c}}\right) ,  \label{e2}
\end{equation}%
where $\theta (x)$ is the Heaviside step function. It is important
to note that the localized spin $S$ of the magnetic ions provides
the main contribution to the DMS magnetization, whereas that of
the itinerant carriers add a minor role. The parameters $M_{0}$
and $C_{0}$ can be easily estimated for $N_{m}$ localized spin
moments leading to the estimation $ M_{0}^{2}/C_{0}
=3SN_{m}/8(S+1)$, which is independent of carriers. Thus, the only
dependence of $F_M$ on the hole population $j$ comes from the term
$T_{c}=T_{c}(j)$ in our approximation.

To obtain a numerical value for $j$, one must also incorporate the
non-magnetic part $F_{N}$ of the free energy for $j$ particles
located in the QD. Unfortunately, the calculation of $F_{N}$
requires the approaches very specific to each individual case as
it depends on the details such as the material composition, size
and shape of the QD, presence of dopants and external fields, etc.
Consequently, this problem cannot be solved in a general manner.
To proceed further, we treat the QD as a scaled QW with a finite
lateral size (and thickness); this is analogous to a nanodot
embedded in a barrier.

The potential profile of the sample structure along the growth
($z$) direction is schematically illustrated in Fig.~1 from the
{\em hole} representation, as is the case throughout the paper. It
is convenient to split $F_{N}$ into two parts,
$F_{N}=E_{j}+F_{1}(T,j)$. The first term,
\begin{equation}
E_{j}=jU+\frac{1}{2}j(j-1)C,  \label{eq2}
\end{equation}%
accounts for the energy acquired by $j$ particles due to their localization
in the QD with the ground state energy $U$ as well as their Coulomb
repulsion energy ($C=e^{2}/\epsilon \sqrt{A_{0}}$, where $e$ is the electron
charge, $\epsilon $ the dielectric constant, and $A_{0}$ the lateral
cross-section of the QD).~\cite{comm01} Then, the remaining part,
\begin{equation}
F_{1}(T,j)=\Omega (T,\mu _{1})+j\mu _{1}(j),  \label{g3}
\end{equation}%
is similar to the free electron gas contribution with the thermodynamic
potential \cite{comm1}%
\begin{equation}
\Omega (T,\mu _{1})=-k_{B}T\sum_{n}\ln [1+e^{(\mu _{1}-\varepsilon
_{n})/k_{B}T}].  \label{e3}
\end{equation}%
$\varepsilon _{n}$ and $\mu _{1}$ represent the energy spectrum (with the
quantum number $n$) and the chemical potential of the QD when the influence
of the magnetic interaction is excluded (i.e., the non-magnetic version of
the QD). Hereinafter, we consider the lateral dimension $A_{0}$ of the QD to
be relatively sizable so that the energy gaps in the discrete energy spectra
are smaller than $k_{B}T$. Then, the sum in Eq.~(\ref{e3}) can be
approximated by an integral with the density of states $mA_{0}/\pi \hbar ^{2}
$. From the relation $j=-\partial \Omega (T,\mu _{1})/\partial \mu _{1}$, $%
\mu _{1}(j)$ is found to be
\begin{equation}
\mu _{1}(j)=k_{B}T\ln \left( e^{\xi j}-1\right) ,  \label{e31}
\end{equation}%
where $\xi =\pi \hbar ^{2}/mA_{0}k_{B}T$ and $m$ is the in-plane
hole effective mass. Subsequently, the thermodynamic potential
[Eq.~(\ref{e3})] can be expressed in the form
\begin{equation}
\Omega \lbrack T,\mu _{1}(j)]=-\frac{k_{B}T}{\xi }\int\limits_{0}^{\infty
}\ln \left[ 1+\left( e^{\xi j}-1\right) e^{-x}\right] dx.  \label{e4}
\end{equation}%
Equations~(\ref{eq2}) and (\ref{g3}) along with $ \mu_{1}(j)$ and
$\Omega (T,\mu _{1})$ from Eqs.~(\ref{e31}) and (\ref{e4})
determine the non-magnetic part of the free energy, while the
total free energy of the DMS QD is the sum $F=F_{M}+F_{N}$.

Now we take into account that the QD is in contact with a large reservoir
providing two-way exchange of carriers through the potential barrier (see
Fig.~1). This leads to the establishment of a unified chemical potential
that coincides with $\mu _{0}$ of the reservoir. Thus, the equation that
determines the population of the QD takes the form
\begin{equation}
\mu _{QD}(j)=\mu _{0}.  \label{e52}
\end{equation}%
Note that $\mu_{QD}(j) \neq \mu _{1}(j)$ since both the
non-magnetic and magnetic interactions are considered for
$\mu_{QD}(j)$. Since the chemical potential $\mu_{QD}(j)$ of the
QD can be expressed as $\mu _{QD}(j)=dF/dj$ in general, the stable
solutions of Eq.~(\ref{e52}) must correspond to the local minima
of $F=F(j)$ or equivalently $d\mu _{QD}(j)/dj>0$.

Finally, the desired solutions require a specific expression for the
dependence $T_{c}=T_{c}(j)$ in Eq.~(\ref{e2}). Following the experimental
data of Ref.~\onlinecite{Boukari}, we adopt a semi-phenomenological
description
\begin{equation}
T_{c}=T_{c}^{0}\left( 1-e^{-\alpha j\xi t}\right) ,  \label{e6}
\end{equation}%
where $T_{c}^{0}$ is the asymptotic value of the critical
temperature (at a sufficiently high hole concentration),
$t=T/T_{c}^{0}$, and $\alpha (=1)$ is the fitting parameter that
adjusts Eq.~(\ref{e6}) to the experiments. Subsequently, the QD
population $j$ can be obtained from Eqs.~(\ref{e2}), (\ref{eq2}),
(\ref{g3}), and (\ref{e52}) as
\begin{eqnarray}
\mu _{0}-U & = &(j-1/2)C+k_B T\ln \left( e^{\xi j}-1\right) -  \nonumber \\
&&k_B T\xi \frac{3SN_{m}}{8(S+1)}e^{-\alpha \xi tj}\left[ 1-\left( \frac{T}{%
T_{c}}\right) ^{2}\right] \theta \left( 1-\frac{T}{T_{c}}\right) .
\label{f7}
\end{eqnarray}

For a numerical evaluation, let us assume the following set of
parameters "typical" for a carrier mediated DMS (e.g., ${\rm
Ga_{0.95}Mn_{0.05}As}$): $m=0.13m_{0}$ ($m_{0}$ is the free
electron mass), $\epsilon =12.9$, $S=5/2$, $N_{m}=1.3\times
10^{21}~ {\rm {{cm}^{-3}}} $$\times$(QD volume), and
$T_{c}^{0}=110$~K. Figure~2 depicts the dependence $\left[ \mu
_{QD}(j)-U\right] /k_{B}T_{c}^{0} $ vs.\ $j$. Clearly, the results
indicate that only one solution for $j$ exists at sufficiently low
or high energies $U$ in reference to $\mu _{0}$ (e.g., dashed line
1 or 3) corresponding to the only stable QD population at a given
$U$. However, the moderate values of $U$ can support multiple
roots. In the case of dashed line 2, two of them (with the
smallest and largest $j$) are stable considering the positive
derivative ($d\mu _{QD}/dj>0$), while the intermediate solution is
not ($d\mu _{QD}/dj<0$). Note also that the stable solutions with
the larger (smaller) $j$ are realized in the FM (PM) phase of the
DMS QD. Hence, this demonstrates a bistable state for a properly
designed QD in terms of the hole population or the magnetic phase.
Figure~3 provides the ranges of $U$ (assuming a fixed $\mu _{0}$)
and $T$ where the bistability can be expected in the system under
consideration.  Clearly, the calculation results indicates a large
window in the system parameter space where the bistability is
possible. The maximum operating temperature may be nearly as high
as $T_{c}^{0}$.

To achieve a bistable state robust against the thermal
fluctuations (i.e., non-volatile) for the possible memory
application, it is necessary to select a condition that provides
the maximal separation $\Delta F=\min \{F_{\max }-F_{P\min
},F_{\max }-F_{F\min }\}$ between the local maximum $F_{\max }$
and each of the local minima $F_{P\min }$ and $F_{F\min }$ (for
the PM and FM phase of the DMS QD, respectively) of the free
energy $F(j)$. At a given temperature, one can find the QD
potential alignment that results in the maximal $\Delta F$.
Figure~4 illustrates the behavior of $F(j)$ at three different
shifts $\Delta \mu =\mu _{0}-U$: curve 1 (3) represents the case
of the monostable state in the FM (PM) state, whereas curve 2
exhibits two local minima in the FM and PM states, respectively,
separated by the maximal $\Delta F$. The temperature dependence of
the maximal $\Delta F$ and the associated optimal potential shift
$\Delta \mu $ is plotted in Fig.~5 along with the mean values of
the particle numbers $j_{P}$, $j_{m}$ and $j_{F}$ corresponding to
$F_{P\min }$,$F_{\max }$ and $F_{F\min }$. Our analysis shows that
the "thermodynamic barrier" $\Delta F$ decreases drastically as $T
$ approaches $T_{c}^{0}$ [curve 2 in Fig.~5(a)].  Obviously, the
system based on the PM-FM transition becomes much less stable near
the critical temperature.

Note that the mean value $j$ is reached through the balance of the particle
flux to and from the QD. Each of these incidents transfers one particle via
the thermally activated processes with a characteristic time $\tau _{0}$,
which depends on the temperature, height and width of the energy barrier
separating the QD and reservoir, etc. The lifetime $T_{lt}$ is defined as
the time it takes to develop a sufficiently large fluctuation to induce a
transition from the state initially at $F_{F\min }=F(j_{F})$ to that at $%
F_{P\min }=F(j_{P})$. Apparently if the system reaches the state at the
local maximum $F_{\max }=F(j_{m})$, further evolution can result in either
the PM (with $j=j_{P}$) or the FM (with $j=j_{F}$) QD state with an
approximately equal probability; hence, $T_{lt}$ can be estimated as the
reciprocal probability for the process $F_{F\min }\rightarrow F_{\max }$ or $%
j_{F}\rightarrow j_{m}$.

The formation of fluctuation $\Delta j=$ $j_{F}-j_{m}$ can be considered
based on the sequential process of hole withdrawal from the QD. For the
first hole transfer out of the QD, the characteristic time of this process
is $\tau _{0}$ as defined above. Due to the finiteness of the QD hole
population, this reduces the chemical potential by $\Delta \mu (1)=\mu
_{QD}(j_{F})-\mu _{QD}(j_{F}-1)$. Consequently, the new time constant
becomes $\tau _{0}\exp [\Delta \mu (1)/k_{B}T]$ when the next hole escapes
from the QD, provided no particles are injected into the QD from the
reservoir. Hence, the mean time necessary for the $\Delta j$ fluctuation
through the sequential withdrawal can be estimated as
\begin{equation}
T_{w}(\Delta j)=\tau _{0}\sum_{j=0}^{j_{m}-1}\exp [\Delta \mu (j)/k_{B}T].
\label{e8}
\end{equation}%
On the other hand, the probability of no hole injection from the reservoir
during this time span $T_{w}$ is $P_{w}=\exp [-T_{w}(\Delta j)/\tau _{0}]$.
The frequency of appearance for such a rare occasion is $P_{w}/\tau _{0}$
and the desired lifetime is
\begin{equation}
T_{lt}=\tau _{0}\exp [T_{w}(\Delta j)/\tau _{0}].  \label{e9}
\end{equation}%
A similar expression is shown to apply to the fluctuations $F_{P\min
}\rightarrow F_{\max }$ or $j_{P}\rightarrow j_{m}$.

One can see that the estimated lifetime (or the bit retention
time) depends crucially on the operating temperature and the QD
sizes, which determine the number of terms $j_{m}$ in the sum of
Eq.~(\ref{e8}). Figure~6 analyzes the results for the GaMnAs QD
with the dimension of $ 25\times 25\times 5$~$\mathrm{{{nm}^{3}}}$
and $15\times 15\times 5$~$ \mathrm{\ {{nm}^{3}}}$ assuming
$T_{c}^{0}=110$~K. As expected, the bistable state becomes less
stable (i.e., shorter $T_{lt}$) as the QD size shrinks. The
thermal fluctuation clearly has a bigger impact in this case due
to the finite number of holes in the QD.   For the two structures
considered (or those of similar sizes), a practically non-volatile
condition (i.e., sufficiently long bit retention) may be achieved
when operating below approx.\ 75 K. $ \protect\tau_{0}=1 $ ns is
used for the calculation.

To examine the feasibility of room temperature application, it is
desirable to extend our consideration of magnetic semiconductors
to those emerging ones with potentially much higher critical
temperatures. Note for instance two recent reports of the DMS with
$T_{c}\geq 300$~K.~\cite{Tsui,Norberg}  Since the search for an
ideal DMS is only at the beginning stage, we assume a hypothetical
material with the characteristics similar to GaMnAs except
$T_{c}^{0}$ in Eq.~(\ref{e6}), which is treated as a variable in a
wide temperature range. Figure~7 depicts the estimated lifetime as
a function of $T_{c}^{0}$, while $T$ is fixed at 300 K.  For a
sufficiently long $T_{lt}$ in this case, the desired material
needs $T_{c}^{0}$ of approx.\ 550 K or higher.

In summary, we investigate theoretically the effect of the reduced
dimensionality and explore the potential bistability conditions
based on the electrically controlled magnetic phase transition in
the magnetic semiconductor nanostructures. The analysis is based
on a semi-phenomenological model that assumes the common magnetic
behavior and a simple hole energy spectrum in a DMS QD.  When
properly designed, the calculation predicts the possibility of
controlled switching between the stable PM and FM states in the
QD.  The parameter window suitable for the bistability formation
is discussed along with the conditions for the maximum
robustness/non-volatility.  An estimation of the bistability
lifetime limited by the thermal fluctuation provides a guideline
for its potential application as a room temperature low-power,
high-density memory element.


This work was supported in part by the Defense Advanced Research
Projects Agency and the SRC/MARCO Center on FENA.

\clearpage

\begin{figure}[tbp]
\includegraphics[scale=.7]{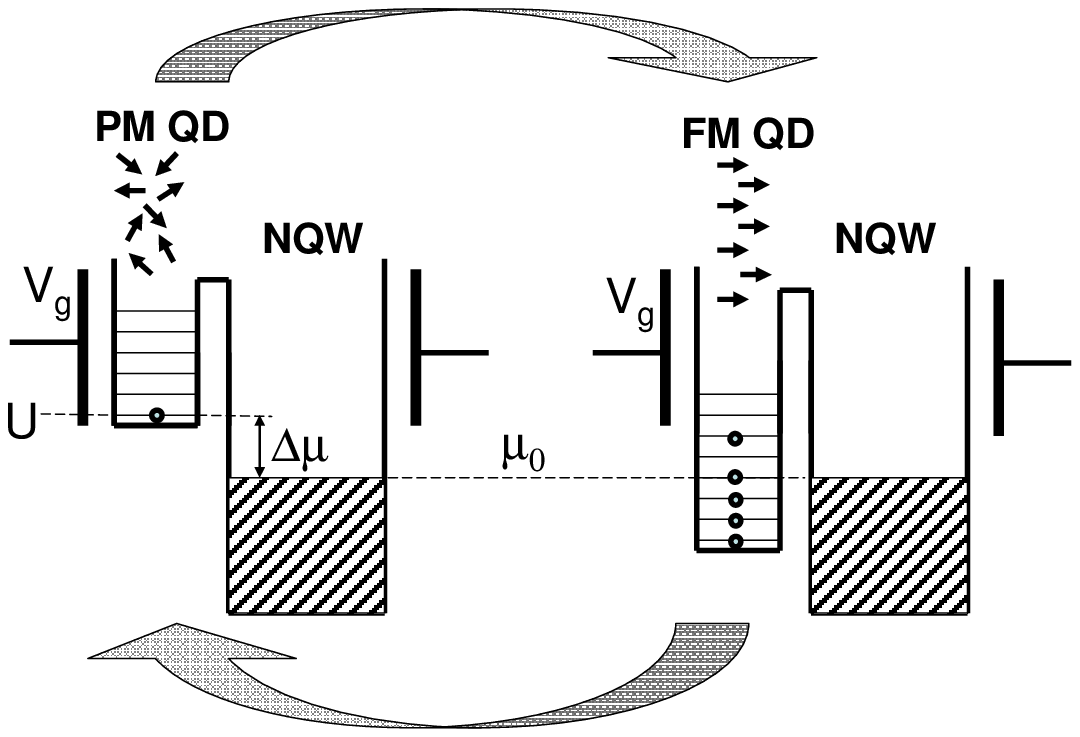}
\caption{ Schematic diagram of the structure containing a DMS QD
(in two different phases) and a nonmagnetic quantum well (NQW)
reservoir separated by a barrier. The energy orientation is
provided from the hole point of view in the valence band. Left:
The PM phase corresponds to disordered magnetic ion spins (small
arrows) and the lack of the magnetic contribution to the hole
energy. This is achieved when the holes (small circles) only
weakly populate the QD with a discrete energy spectrum and, thus,
cannot change its magnetic state.  The ground state $U$ of the QD
is relatively high in reference to the chemical potential
$\protect\mu_{0}$ of the hole reservoir. Right: Another
thermodynamically stable state (at the same external conditions)
is possible when the magnetic ions are ferromagnetically ordered.
Magnetic interactions can decrease the hole potential that the
ground state of the DMS QD is now sufficiently below
$\protect\mu_{0}$; the equilibrium hole population is high enough
to stabilize the FM phase. Switching between the PM and FM states
can be achieved by applying a gate bias $V_{g}$ that populates or
depopulates the DMS QD.}
\end{figure}

\clearpage
\begin{figure}[tbp]
\vspace{1.3cm}
\includegraphics[scale=.8]{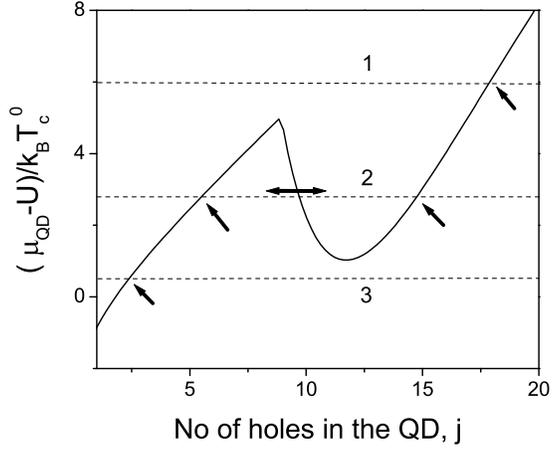}
\caption{Chemical potential of the DMS QD with the thickness of 5
nm and the cross-section of $25\times 25$~$\mathrm{{{nm}^{2}}}$ as
a function of the QD hole population [Eq.~(\protect \ref{f7})].
The parameters of ${\rm Ga_{0.95}Mn_{0.05}As}$ are assumed with
$T=$ 77 K as discussed in the text.  The solutions of
Eq.~(\protect\ref{f7}) can be found as intersections of the solid
curve with the horizontal line corresponding to a certain value of
$\Delta\protect\mu$ (=$\mu _{0}-U$). Two cases $
\Delta\protect\mu/T_{c}^{0}=$6 and 0.5 (dashed lines 1 and 3)
provide monostable FM and PM states, while dashed line 2
($\Delta\protect\mu/T_{c}^{0}=2.7 $) depicts the bistable state.
Stable solutions are indicated by single-head arrows and the
unstable one by the horizontal double-head arrow. }
\end{figure}

\clearpage

\begin{figure}[tbp]
\vspace{1.3cm}
\includegraphics[scale=.8]{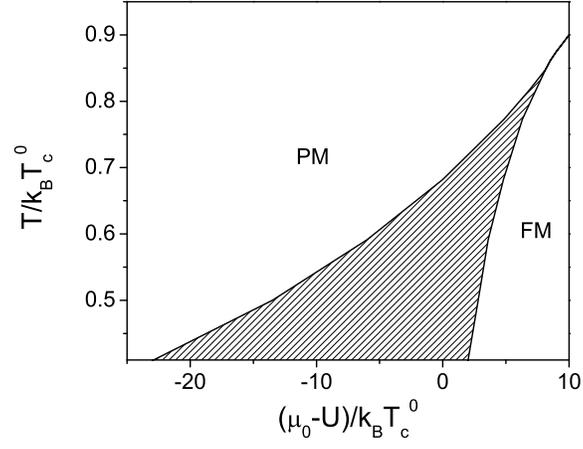}
\caption{Phase diagram of the parameter space indicating the
potential bistability region (the shaded area).  The PM and FM
denote the monostable areas corresponding to the PM and FM QD
states, respectively.  The same parameters as in Fig.~2 are
assumed ($T_{c}^{0}=$ 110 K).}
\end{figure}

\clearpage

\begin{figure}[tbp]
\vspace{1.3cm}
\includegraphics[scale=0.8]{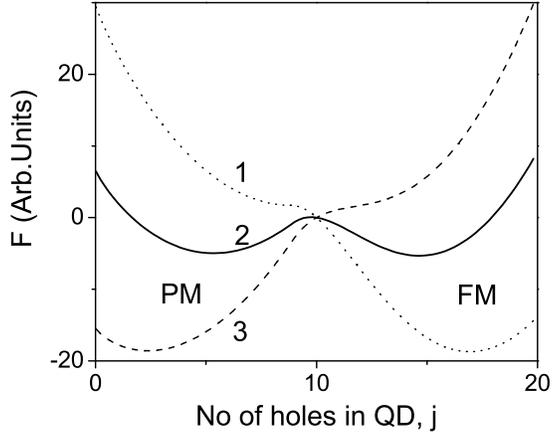}
\caption{Free energy of the QD calculated as a function of hole
population for three different values of $\Delta\protect\mu$
(=$\mu _{0}-U$): $\Delta\protect\mu/T_{c}^{0}$=6 (curve 1, dotted
line); 2.7 (curve 2, solid line); 0.5 (curve 3, dashed line). The
single minima of curves 1 (FM phase) and 3 (PM phase) correspond
to the vicinities of the right and left boundaries of the bistable
area in Fig.~3; curve 2 represent the bistable case with the
optimal free energy barrier height separating two local minima.
The same parameters as in Fig.~2 are assumed ($T=$ 77 K,
$T_{c}^{0}=$ 110 K).}
\end{figure}

\clearpage
\begin{figure}[tbp]
\vspace{1.3cm}
\includegraphics[scale=0.5]{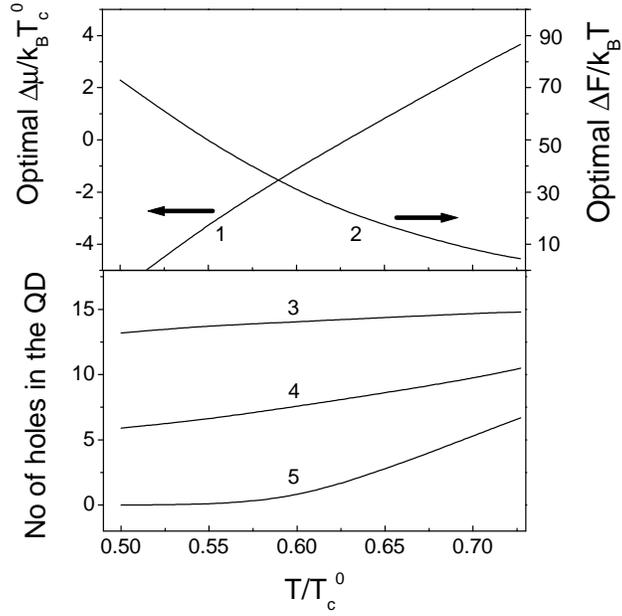}
\caption{Maximal free energy barrier height (in units of $k_{B}T$)
between the local maximum and the closest local minimum in the
bistable case (curve 2) and the corresponding optimal chemical
potential (curve 1) as a functions of temperature. Curves 3$-$5 in
the bottom pane shows the hole population at the FM minimum (curve
3), the PM minimum (curve 5), and the local maximum (curve 4) of
$F(j)$ for the bistable case shown in the top pane at the
corresponding temperature. The same parameters as in Fig.~2 are
assumed ($T_{c}^{0}=$ 110 K)}
\end{figure}

\clearpage

\begin{figure}[tbp]
\vspace{1.3cm}
\includegraphics[scale=.8]{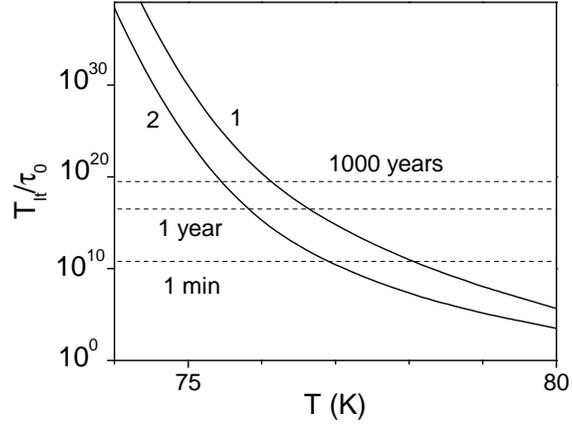}
\caption{Bistability lifetime vs. temperature at $T_{c}^{0}=110$
K.  Two different QD dimensions are considered: (1) $25\times
25\times 5$~$\mathrm{{{nm}^{3}}}$ and (2) $15\times 15\times 5
$~$\mathrm{{{nm}^{3}}}$.  The mean time $ \protect\tau_{0}$ of
particle exchange between the QD and reservoir via thermal
processes is assumed to be 1 ns.  Other parameters are the same as
in Fig.~2.}
\end{figure}

\clearpage
\begin{figure}[tbp]
\vspace{1.3cm}
\includegraphics[scale=.8]{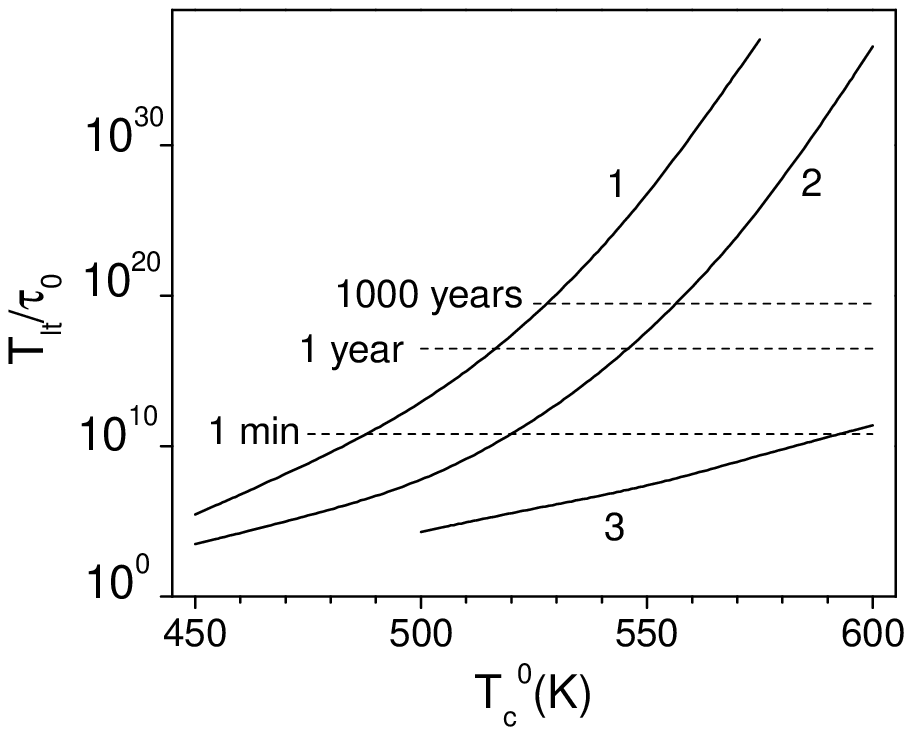}
\caption{Bistability lifetime vs. temperature for a hypothetical
DMS material with the characteristics similar to GaMnAs except
$T_{c}^{0}$ in Eq.~(\ref{e6}), which is treated as a variable in a
wide temperature range.  $T$ is fixed at 300 K and three different
QD sizes are considered: (1) $25\times 25\times 5
$~$\mathrm{{{nm}^{3}}}$; (2) $ 15\times 15\times
5$~$\mathrm{{{nm}^{3}}}$; and (3) $25\times 25\times 3$~$
\mathrm{{{nm}^{3}}}$. The mean time $ \protect\tau_{0}$ of
particle exchange between the QD and reservoir via thermal
processes is assumed to be 1 ns as in Fig.~6}
\end{figure}


\begin{thebibliography}{99}
\bibitem{comm1} Here we go from the canonical to the grand canonical Gibbs ensemble
treating $j$ as the mean value $\left\langle j\right\rangle $ over
the ensemble.
\end{thebibliography}
\end{document}